\documentclass[%
showpacs,
 amsmath,amssymb,
 aps,
twocolumn,
eqsecnum,
]{revtex4}

\usepackage{epsfig}
\usepackage{euscript,graphicx}
\usepackage{amsthm,dsfont,amsfonts,amsmath,amssymb}
\usepackage{euscript,color,fontenc,textcomp,relsize}
\usepackage{bm,url,float,cleveref}
\usepackage[caption=false]{subfig}
\usepackage[english]{babel}
\usepackage{natbib}

\allowdisplaybreaks
\newcommand{\vek}[1]{\bm{#1}}
\newcommand{\vL}{\vek{L}}
\newcommand{\vS}{\vek{S}_1}
\newcommand{\vSS}{\vek{S}_2}
\newcommand{\no}{\\\notag}
\begin{document}

\preprint{APS/123-QED}

\title{Analytic Keplerian-type parametrization for general spinning compact binaries with the leading order spin-orbit interaction}

\author{ Gihyuk Cho\,$^{1,2}$\footnote{whrlsos@snu.ac.kr},
 Hyung Mok Lee\,$^2$}

\affiliation{
$^{1}$
Department of Physics and Astronomy,
Seoul National University,
Seoul 151-742, Korea\\
$^2$ Korea Astronomy and Space Science Institute,
776 Daedeokdaero, Daejon, Korea
}


\date{\today}

 \begin{abstract}
We derive a fully analytic Keplerian-type parametrization solution to conservative motion of spinning binary in ADM gauge. This solution is able to describe three dimensional motion of binaries of arbitrary eccentricity, mass ratio and initial configuration of spin angular momentum up to the leading order of post-Newtonian(PN) approximation and  a linear order in spin. Based on our results waveforms can be quickly computed with high accuracy.
\end{abstract}

\maketitle

%
%
\section{\label{sec:level1}Introduction}
Under Newtonian gravity which has only mass as a source, the two body problem can be analytically solved into the well-known Kepler parametrized solution. But under the Einstein gravity which has a stress energy tensor of 10 components as its source, parameters more than mass must be considered. In the classical works \cite{Mathisson2010, Corinaldesi1951,Dixon1970,Bailey1983}, multipolar expansion of stress energy tensor on a generic background metric field, has been made under the assumption that matter is highly localized in space. The results, which involve mass as a monopole contribution while spin as a dipole moment, give us understanding on how masses and spin angular momenta of point particles affect the entire dynamics. Unlike Newtonian gravity, spin angular momentum produces extra gravitational effects and thus makes orbit plane of two body and spin axes precess.\\ 
\indent Even though dynamics of binary systems in general relativity can be solved by numerical relativity (NR) with high accuracy, it is still worth solving the equation of motion analytically. Actually, the post-Newtonian (PN) approximation which has been well established \cite{Blanchet2014,Bernard2018}, provides spin precession equation \cite{A.Gergely1998,Gergely2000,Blanchet2006,Faye2006} and the orbital evolution equations up to at least 4PN order \cite{Damour2014,Bernard2018}. Although the PN equations can be integrated numerically with much smaller amount of computer resources than NR but analytical solution would be desirable for quick parameter estimation \cite{Veitch2015,Farr2016}. Until now analytic solutions have been obtained only for some limited cases such as quasi-circular, nearly aligned\cite{Klein2013}, slowly spinning \cite{Chatziioannou2013a} or almost equal mass\cite{Konigsdorffer2005,Tessmer2009}. General cases with arbitrary spin can be solved up to a certain PN order in terms of elliptic functions \cite{Racine2008} and recent works present a form of the solutions partially in terms of sine-like Jacobi elliptic function with integrating out the oscillating part of orbital motion  \cite{Marsat2014,Chatziioannou2017}.
In this paper, we present an analytic parametrized solution which involves both spin precession and orbital evolution for the binaries with arbitrary spins, masses and eccentricities of orbit. Namely, our solution parameterizes tilt angle of orbital planes and precession of spin angular momentum with respect to a fixed inertial frame, radial and angular motion of relative displacement of binaries and physical time evolution only via the eccentric anomaly, a Keplerian parameter that serves as an independent variable. Additionally because we express the entire parameterization in closed form witout any approximations, the solution is fully general.\\ 
\indent This paper is organized as follows. In \S\ref{eom}, dynamics which we are going to solve is presented in terms of Hamiltonian. The Hamiltonian includes Newtonian order of orbital dynamics and the leading order of spin-orbit interaction term (linear in spin). Parameters and reference frames which we choose are also presented.  In \S\ref{solution}, the ways to get the Keplerian-type parametrizations and their results are given in the following orders : Relative motion of angles between angular momenta (\S\ref{solutionAngle}),  the absolute precession of the orbital angular momentum in an inertial frame (\S\ref{solutionOAM}), and finally, the relative position of compact binaries (\S\ref{solutionRP}). In \S\ref{almostequalmass}, we provide an approximant for almost equal mass case and compare it with the previous work \cite{Konigsdorffer2005} as a sanity check. The final secion summarized our results.


\section{\label{eom} Dynamics with leading order of spin-orbit interaction}
\subsection{The Hamiltonian}
In \textit{Arnowitt-Deser-Misner}(ADM) type cooridnate \cite{Arnowitt2004} and with the \textit{Newton-Wigner-Pryce} supplementary spin condition \cite{Newton1949,Pryce1948}, the (reduced) Hamiltonian $H$ for compact bianary systems composed of two stars with masses $m_1$ and $m_2$, is as follows \cite{Wex1995},
\begin{align}\label{Hamiltonian}
H =\frac{\vek{p}^2}{2}-\frac{1}{r}+\frac{1}{c^2r^3}\vek{L}\cdot\vek{S}_\text{eff}\,,
\end{align}
where $c$ is the speed of light, $\vek{r}=\frac{\vek{R}}{G\,(m_1+m_2)}$, $\vek{p}=\frac{\vek{\mathcal{P}}\,(m_1+m_2)^2}{m_1\,m_2}$ with $\vek{R}$ and $\vek{\mathcal{P}}$ being relative position vector and conjugate momentum, respectively. The orbital angular momentum $\vek{L}$ and effective spin $S_\text{eff}$ are defined as
\begin{align}
\vek{L}=\vek{r}\times\vek{p}\,,
\end{align}
and
\begin{align}
\vek{S}_\text{eff} = \delta_1\,\vek{S}_1 +\delta_2\, \vek{S}_2\,.
\end{align}
Here $\vS$ and $\vSS$ are spin vectors of $m_1$ and $m_2$ and the waiting factors of $\delta_1$ and $\delta_2$ are,
\begin{subequations}
\begin{align}
\delta_1 = 2\,\frac{m_1\,m_2}{M^2}\left(1+\frac{3}{4}\frac{m_2}{m_1}\right)\,,\\
\delta_2 = 2\,\frac{m_1\,m_2}{M^2}\left(1+\frac{3}{4}\frac{m_1}{m_2}\right)\,.
\end{align}
\end{subequations}
For a consistent post-Newtonian approximation, we need to include the first correction of orbital dynamics (1PN) but intentionally exclude it because 1PN correction to orbital motion does not affect the spin precession at the leading order and bringing the 1PN correction into the result is expected to be straightforward \cite{Konigsdorffer2005,Damour1985}.

\subsection{The equation of motion}
\begin{figure}
										 \includegraphics[width=0.56\textwidth]{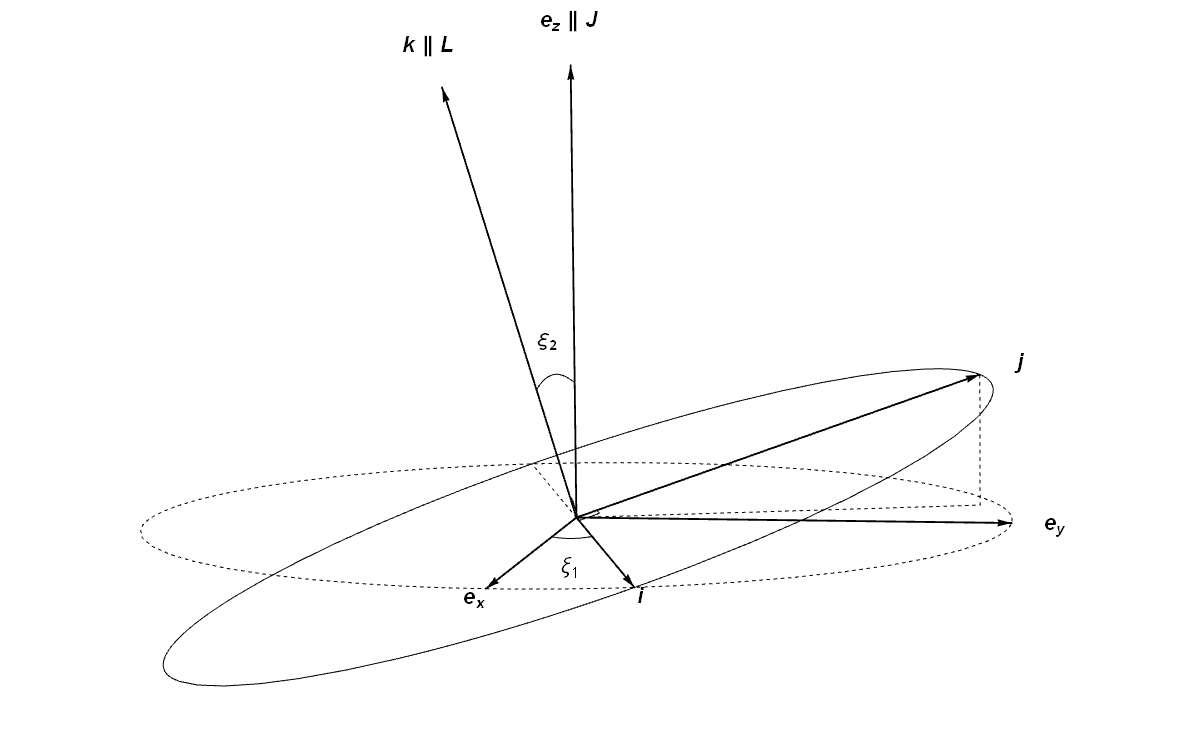}
\centering
\caption{Two bases are displayed : The inertial frame $(\vek{e}_x,\,\vek{e}_y,\,\vek{e}_z)$ and the non-inertial frame $(\vek{i},\,\vek{j},\,\vek{k})$. The total angular momentum $\vek{J}$ is parallel to $\vek{e}_z$, and the orbital angular momentum $\vek{L}$ to $\vek{k}$. The non-inertial frame is constructed by rotating the inertial frame around $z$-axis by $\xi_1$ and then around $x$-axis by $\xi_2$, so that  $\xi_1$ and $\xi_2$ determine the orientation of $\vek{L}$.  }
    \label{fig:frames}
\end{figure}

The equation of motion is given by Poisson bracket,
\begin{align}
\dot{\vek{r}}:=\frac{d\vek{r}}{dt}=\{\vek{r},\,H\}\,.
\end{align}
Likewise the corresponding angular momentum and spin precession equations are given as
\begin{subequations}\label{spinp}
\begin{align}
\label{lEq}\frac{d\vek{L}}{dt}&=\{\vek{L},\,H\}=\frac{1}{c^2\,r^3}\,\vek{S}_\text{eff}\times\vek{L}\,,\\
\frac{d\vek{S}_1}{dt}&=\{\vek{S}_1,\,H\}=\frac{\delta_1}{c^2\,r^3}\,\vek{L}\times\vek{S}_1\,,\\
\frac{d\vek{S}_2}{dt}&=\{\vek{S}_2,\,H\}=\frac{\delta_2}{c^2\,r^3}\,\vek{L}\times\vek{S}_2\,,
\end{align}
\end{subequations}
by invoking the fact that angular momentum is a generator of rotations \textit{i.e.}
\begin{align}
\{S_i,S_j\} =\epsilon_{ijk}\,S_k\,.
\end{align}
From the evolution of angular momentum and spins, we can find several conserved quantities that can be exploited in deriving the dynamical solutions. First, the magnitudes of the angular momenta are conserved because
\begin{subequations}
\begin{align}
\frac{d (\vek{L}\cdot\vek{L})}{dt}=\frac{2}{c^2\,r^3}\,\vek{L}\cdot(\vek{S}_\text{eff}\times\vek{L})=0\,,\\
\frac{d (\vek{S}_1\cdot\vek{S}_1)}{dt}=\frac{2\,\delta_1}{c^2\,r^3}\,\vek{S}_1\cdot(\vek{L}\times\vek{S}_1)=0\,,\\
\frac{d (\vek{S}_2\cdot\vek{S}_2)}{dt}=\frac{2\,\delta_2}{c^2\,r^3}\,\vek{S}_2\cdot(\vek{L}\times\vek{S}_2)=0\,.
\end{align}
\end{subequations}
Also in the absence of the angular momentum loss from the binary system, the \textit{total angular momentum} $\vek{J} =\vL+\vS+\vSS$ is conserved.
Without loss of generality we can define the z-axis of the inertial frame to be aligned with $\vek{J}$ while the the orthogonal axes $x$ and $y$ can be chosen arbitrarily on the plane perpendicular to $z$-axis as shown in Fig.~\ref{fig:frames}.

\begin{figure}[H]
	\centering
	\includegraphics[width=0.49\textwidth]{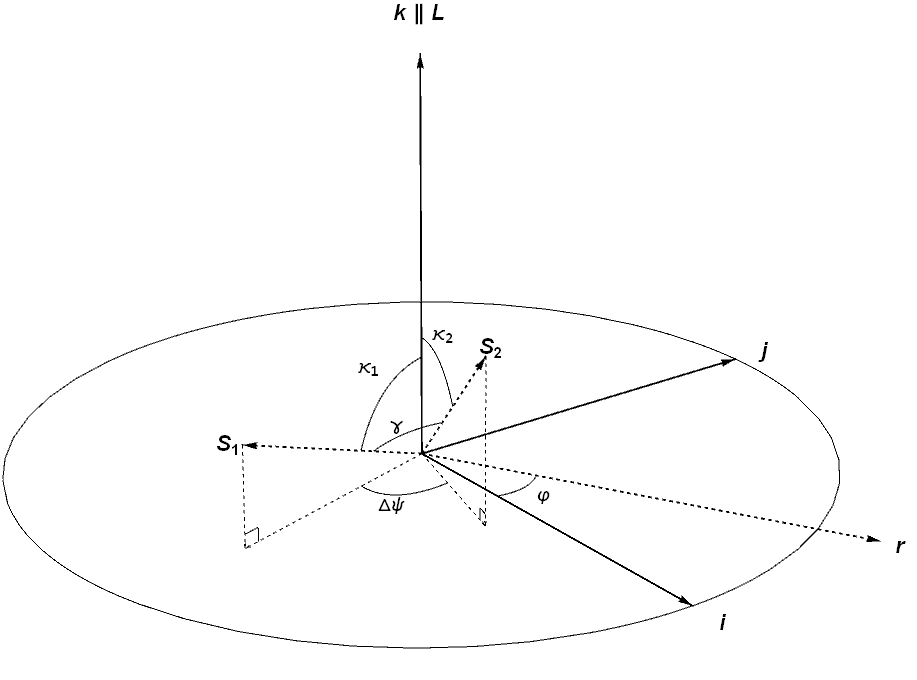}
    \caption{The geometry of spin and angular momentum vectors, and the inertial frame.}
    \label{fig:angles}
\end{figure}

At the same time, we define an orthonormal non inertial frame $(\vek{i},\vek{j},\vek{k})$,

\begin{align}
\left(\begin{array}{ccc}
 \vek{i}\\
 \vek{j}\\
 \vek{k}\\
\end{array}\right)=
\Lambda\,
\left(\begin{array}{ccc}
 \vek{e}_x\\
 \vek{e}_y\\
 \vek{e}_z\\
\end{array}\right)
\end{align}
where $\Lambda$ is the Euler matrix as shown below
\begin{align}
\Lambda=\left(
\begin{array}{ccc}
 \cos\xi_1& \sin\xi_1 &0\\
 -\sin\xi_1\,\cos\xi_2& \cos \xi_1\,\cos\xi_2 &\sin\xi_2\\
 \sin\xi_1\sin\xi_2& -\cos\xi_1\sin\xi_2 & \cos\xi_2\\
\end{array}
\right)\,.
\end{align}
The geometric meaning of the Euler matrix is presented in Fig.~\ref{fig:frames}.\\
\indent Next, we define the angles $\gamma$, $\kappa_1$ and $\kappa_2$ in the range of  $[0,\, \pi]$ such that
\begin{subequations}\label{angles}
\begin{align}
\cos\gamma := \frac{\vek{S}_1\cdot\vek{S}_2}{S_1\, \,S_2}\,,\\
\cos\kappa_1 :=\frac{\vek{L}\cdot\vek{S}_1}{L\,\, S_1}\,,\\
\cos\kappa_2 :=\frac{\vek{L}\cdot\vek{S}_2}{L\, \,S_2}\,,
\end{align}
\end{subequations}
as illustrated in Fig.~\ref{fig:angles}.
The rates of changes of these angles are given via Eq.(\ref{spinp}) \textit{i.e.}
\begin{subequations}\label{anglesEOM}
\begin{align}
\frac{d \cos\gamma}{dt}& =\,\frac{\delta_1-\delta_2}{c^2\,r^3 \,S_1\,S_2}\vek{L}\cdot(\vek{S}_1\times\vek{S}_2)\, ,\\
\label{anglesEOM2}\frac{d \cos\kappa_1}{dt}& =\,\frac{\delta_2}{c^2\,r^3 \,S_1\,L}\vek{L}\cdot(\vek{S}_1\times\vek{S}_2)\,,\\
\frac{d \cos\kappa_2}{dt}& =\,-\frac{\delta_1}{c^2\,r^3 \,S_2\,L}\vek{L}\cdot(\vek{S}_1\times\vek{S}_2)\,.
\end{align}
\end{subequations}
By observing their common factor $\vek{L}\cdot(\vek{S}_1\times\vek{S}_2)$, we can define new conserved quantities $\sigma_1$, $\sigma_2$ as follows
\begin{subequations}\label{sigmas}
\begin{align}
\sigma_1 \,&:= \cos\gamma-\frac{L}{S_2}\frac{\delta_1-\delta_2}{\delta_2}\cos\kappa_1\,,\\
\sigma_2\,&:= \cos\kappa_2+\frac{\delta_1}{\delta_2}\frac{S_1}{S_2}\cos\kappa_1\,.
\end{align}
\end{subequations}
Note that if $\delta_1-\delta_2=0$, or equal mass case, then $\sigma_1=\cos\gamma$ implying that the angle $\gamma$ between $\vS$ and $\vSS$ is conserved.\\
\indent Through straightforward vector algebra, one can show that $\vek{L}\cdot(\vek{S}_1\times\vek{S}_2) = L\,S_1\,S_2\,\sin\kappa_1\sin\kappa_2\sin\Delta\psi$, where $\Delta\psi$ is an angle between the projection of $\vS$ and $\vSS$ onto x-y plane of the non-inertial frame where $\vL$ is along with $z$-axis (See Fig.~\ref{fig:angles}). It has the following relation with the others,
\begin{align}\label{cosdeltapsi}
\cos\Delta\psi=\frac{\cos\gamma-\cos\kappa_1\,\cos\kappa_2}{\sin\kappa_1\,\sin\kappa_2}\,.
\end{align}
When either $\sin\kappa_1=0$ or $\sin\kappa_2=0$, $\Delta\psi$ is not defined in the geometric sense. But physically we can extend the definition of $\Delta\psi$ properly. For example let us assume that $\kappa_1=0$ and $\sin\kappa_2\neq0$ at the initial time $t_0$. Then, during an infinitesimal time interval $dt$, $d\vS=0$ because $|\frac{d\vS}{dt}|\sim|\vL\times\vS|\sim\sin\kappa_1=0$. On the other hand, $d\vL=\vL(t_0+dt)-\vL(t_0)=-\frac{dt\,\delta_2}{c^2\,r^3}\vL\times\vSS$. Therefore the square of the $d\vL$ is given as
\begin{align}\label{dL}
\big|\vL(t_0+dt)-\vL(t_0)\big|^2=dt^{\,2}\,\left(\frac{L\,S_2\,\delta_2\,\sin\kappa_2}{c^2\,r^3}\right)^2\,.
\end{align}\\
Because $\vL(t_0)$ and $\vS(t_0)$ are aligned and $\vS(t_0+dt)=\vS(t_0)$, the infinitesimal change of $\kappa_1$ is the angle between $\vL(t_0+dt)$ and $\vL(t_0)$\,. Thus the right hand side of Eq.(\ref{dL}) is
\begin{align}
    2\,L^2\,(1-\cos d\kappa_1)
    =L^2\,d\kappa_1^{\,2}\,+O(d\kappa_1^{\,3}).
\end{align}
Since $\kappa_1$ can only increase during $dt$, we get $\frac{d\kappa_1}{dt}=\frac{\delta_2\,S_2\,\sin\kappa_2}{c^2\,r^3}$. Thus by setting $\sin\Delta\psi=-1$, replacing the vectorial expression $\vL\cdot(\vS\times\vSS)$ in Eqs.(\ref{anglesEOM}) to the algebraic expression $L\,S_1\,S_2\,\sin\kappa_1\sin\kappa_2\sin\Delta\psi$ becomes valid even in the case of $\kappa_1=0$. Likewise we arrive at same extension $\sin\Delta\psi=-1$ when $\kappa_2=\pi$ (with $\sin\kappa_1\neq0$), while $\sin\Delta\psi=1$ in the case of $\kappa_1=\pi$ ($\sin\kappa_2\neq0$) and $\kappa_2=0$ ($\sin\kappa_1\neq0$). Mathematically, this kind of extension is nothing but taking the limit of $\sin\kappa_{1,2}\rightarrow0$ along the physically preferred paths. If $\sin\kappa_1=0$ and $\sin\kappa_2=0$ are both satisfied at some point, since every angle becomes constant, we do not need to define $\Delta\psi$.\\ 

We have three time dependent variables $\kappa_1(t)$, $\kappa_2(t)$ and $\gamma(t)$ to solve and two conserved quantities, $\sigma_1$ and $\sigma_2$. So, if we determine a numerical value of any one angle (say $\kappa_1$), the others are determined automatically by the initially fixed values of $\sigma_1$ and $\sigma_2$. This fact allows the relations of Eqs.(\ref{anglesEOM}), where $\kappa_1$, $\kappa_2$ and $\gamma$ are non-linearly entangled, to be expressed in a single variable equation, for example $\kappa_1$. This specific choice of $\kappa_1$ is followed from the fact that $\gamma$ is not a good parameter. As mentioned earlier, $\gamma$ fails to parametrize the dynamics when mass ratio becomes close to 1. Choosing $\gamma$ as a paramter was used in \cite{Kesden2015}, and we can see that there are divergences as $q\rightarrow1$ in the equations (7a) and (7b) in \cite{Kesden2015}. 

\section{The Keplerian-type parametrization}\label{solution}
\subsection{The Keplerian-type parametrization of the Angles}\label{solutionAngle}
\begin{widetext}
In this section, we solve the Eq.(\ref{anglesEOM})s. We reduce the common factor $\vek{L}\cdot(\vek{S}_1\times\vek{S}_2)$, using $\sigma_1$ and $\sigma_2$, into single variable ($\cos\kappa_1$) dependent expression,
\begin{align}\label{LS1S2}
&\frac{\vek{L}\cdot(\vek{S}_1\times\vek{S}_2)}{L\,S_1\,S_2 } = \sin\kappa_1\sin\kappa_2\sin\Delta\psi\,,\\\notag
= &\,\pm\sqrt{1-\cos^2\kappa_1-\cos^2\kappa_2-\cos^2\gamma+2\cos\gamma\cos\kappa_1\cos\kappa_2}
\,\\\notag
=&\,\frac{\pm}{\delta_2\, S_2}\bigg\{-2\, \delta_1 \,L \,S_1 (\delta_1-\delta_2) \cos ^3\kappa_1-\big[L^2\, (\delta_1-\delta_2)^2+2\, \delta_2\, L \,\sigma_2\, S_2 \,(\delta_2-\delta_1)+\delta_1^2\, S_1^2+2 \,\delta_1\, \delta_2 \,\sigma_1\, S_1\, S_2+\delta_2^2 \,S_2^2\big]\,\cos^2\kappa_1 \\\notag
&+2 \,\delta_2\, S_2\, \big[L\,\sigma_1 \,(\delta_2-\delta_1)+\sigma_2\, (\delta_1 S_1+\delta_2 \sigma_1 S_2)\big]\,\cos \kappa_1 -\delta_2^2\, S_2^2 \,(\sigma_1^2+\sigma_2^2-1)\bigg\}^{1/2}\,.
\end{align}
The sign of $\vek{L}\cdot(\vek{S}_1\times\vek{S}_2)$ is determined by the sign of $\sin\Delta\psi$. For simplicity, let us denote $\cos\kappa_1=x$ and factorize the expression inside the square root in terms of its 3rd polynomial roots of $x=(x_1,x_2,x_3)$ and introduce a coefficient $A:=2\,L\,S_1\,\delta_1\,(\delta_2-\delta_1)$ to simplify the above equation as
\begin{align}
\frac{\vek{L}\cdot(\vek{S}_1\times\vek{S}_2)}{L\,S_1\,S_2 } =\,\frac{\pm}{\delta_2\, S_2}\sqrt{A\,(x-x_1)(x-x_2)(x-x_3)}\,.
\end{align}
\end{widetext}

Before proceeding integrations, one needs to clarify the existence of the real roots of $x$. Here is a brief proof: Let us assume that there is no instance such that $\kappa_1(t)$, $\kappa_2(t)$ and $\gamma(t)$ satisfy $\vL\cdot(\vS\times\vSS)=0$ during evolutions. This assumption says that if $\vL\cdot(\vS\times\vSS)>0$ initially this inequalitiy always holds by virtue of continuity. Also the followings hold forever,  $\frac{d\cos\kappa_1}{dt}>0$, $\frac{d\cos\gamma}{dt}>0$ (if $\delta_1>\delta_2$) and $\frac{d\cos\kappa_2}{dt}<0$. Then we are able to find a real value $T$ such that $\cos\kappa_1>0$, $\cos\gamma>0$ and $\cos\kappa_2<0$ at times $t>T$ unless $\frac{d\cos\kappa_1}{dt}\rightarrow0+$, $\frac{d\cos\gamma}{dt}\rightarrow0+$ and $\frac{d\cos\kappa_2}{dt}\rightarrow0-$ fast enough as $t\rightarrow\infty$. But we exclude this dissipative possibility because governing dynamics is symmetric under the reflection of $t\rightarrow -t$. \\
Meanwhile the time derivative of $\vL\cdot(\vS\times\vSS)$ is
\begin{align}
    &\frac{d\vL\cdot(\vS\times\vSS)}{dt}\\\notag
    =&\frac{-2\cos\kappa_1+2\cos\gamma\,\cos\kappa_2}{\vL\cdot(\vS\times\vSS)}\frac{d\cos\kappa_1}{dt}\\\notag
    +&\frac{-2\cos\kappa_2+2\cos\gamma\,\cos\kappa_1}{\vL\cdot(\vS\times\vSS)}\frac{d\cos\kappa_2}{dt}\\\notag
    +&\frac{-2\cos\gamma+2\cos\kappa_1\,\cos\kappa_2}{\vL\cdot(\vS\times\vSS)}\frac{d\cos\gamma}{dt}\,.
\end{align}
It is obvious that $\frac{d\vL\cdot(\vS\times\vSS)}{dt}<0$ when $t>T$. Similarly we can estimate its second time derivative and we can decompose it into two components \textit{i.e.} $\frac{d^2\vL\cdot(\vS\times\vSS)}{dt^2}=\hat{A}(t)+\hat{B}(t)\,\dot{r}$ where $\hat{A}(t)<0$, $\hat{B}(t)>0$ but $\dot{r}$ oscillates sinusoidally between positive and negative values with zero averaged value \textit{i.e.} $\frac{1}{2\pi}\int_0^{2\pi}du\,\dot{r}=0$. Thus in much longer time scale than one orbital period, the absolute value of $\frac{d\vL\cdot(\vS\times\vSS)}{dt}$ increases, \textit{i.e.} decreasing of $\vL\cdot(\vS\times\vSS)$ accelerates. This suggests that at some point $\vL\cdot(\vS\times\vSS)$ will pass zero, or $\vL\cdot(\vS\times\vSS)>0$ cannot hold forever. This is contradictory to the assumption. Likewise another assumption that $\vL\cdot(\vS\times\vSS)<0$ also encounters the same contradiction.  Finally we come to the conclusion that whatever the initial condition is, binary systems eventually meet the configuration which corresponds to $\vL\cdot(\vS\times\vSS)=0$. Actually this happens twice, once when $\vL\cdot(\vS\times\vSS)$ is increasing or once when it is decreasing. Let us match cosine value of these two $\kappa_1$ values to $x_2$ and $x_3$. Then it is obvious that they are real and the absolute values are less than or equal to $1$. Additionally, from the fact that cubic polynomicals of which all coefficients are real, cannot have two real roots and a single complex root, we can conclude that $x_1$ is also real. Note that our assumption does not include the equal mass case, \textit{i.e.}, $\delta_1=\delta_2$ in the above proof. Since there is no reason for any discontinuity toward the equal mass case, we can extend the validity of our formalism to $\delta_1=\delta_2$ although we need a special caution in applying the formalism presented below when two masses become very close. The case of nearly equal mass is treated separately in \S\ref{almostequalmass}.

\begin{figure}[H]
	\centering
	\includegraphics[width=0.4\textwidth]{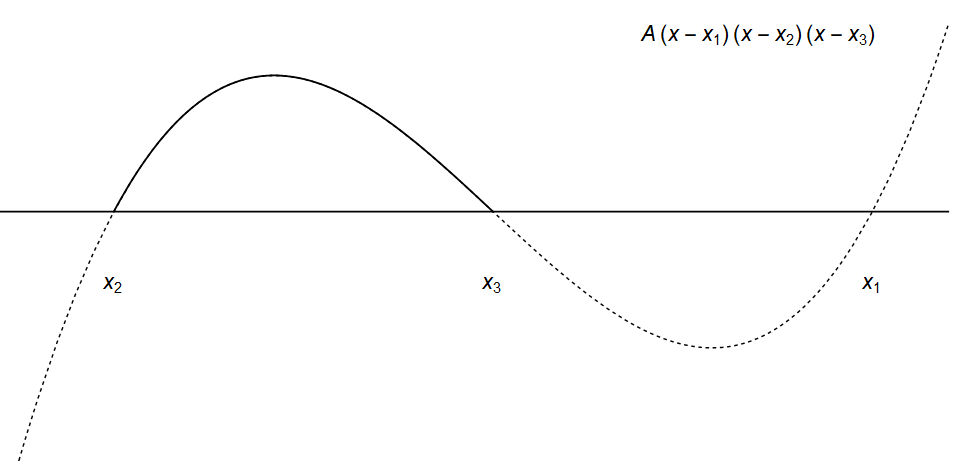}
	\caption{The shape of the function inside the square-root of Eq.(\ref{LS1S2}).}
    \label{fig:root}
\end{figure}

\indent Without loss of generality, we assume that $A> 0$ and $x_2\leq\cos\kappa_1\leq x_3$, because it must be non negative within the square root as illustrated in Fig.~\ref{fig:root}.
Eq.(\ref{anglesEOM2}) becomes,
\begin{align}
\frac{d x}{dt}& =\frac{\pm}{c^2\,r^3}\sqrt{A(x-x_1)(x-x_2)(x-x_3)}\,,
\end{align}
or,
\begin{align}\label{ck1Eq}
\frac{ \, d\cos\kappa_1}{\sqrt{A(\cos\kappa_1-x_1)(\cos\kappa_1-x_2)(\cos\kappa_1-x_3)}} = \frac{\pm\,dt}{c^2\,r^3}.
\end{align}
It is enough that $r$ is given in the Newtonian order, because we restrict the spin orbit interaction only up to the leading order. In Newtonian order we can use the well known expressions of $r = a\,(1-e\,\cos u) $ and $n \,(t - t_0) = u-e\,\sin u$, where $u$ is the eccentric anomaly, $n$ is the mean angular speed and $t$ is physical time with an initial value $t_0$ when $u=0$, $a$ is the semi-major axis of the ellipse, and $e$ is the eccentricity. Thus the right hand side of Eq.(\ref{ck1Eq}) can be integrated to
\begin{align}\label{rhs}\notag
(r.h.s) =&\int du\frac{dt}{du}\frac{\pm1}{c^2\,r^3}\\\notag
=&\alpha \pm \frac{\nu+e\sin\nu}{c^2\,n \,a^3 \,(1-e^2)^{3/2}}\,,\\
=&\alpha \pm \frac{\nu+e\sin\nu}{c^2\,L^3}\,+O(\frac{1}{c^3}),
\end{align}
where we use the Newtonian expression of the magnitude of the orbital angular momentum $L=n^\frac{1}{3}\,a\,(1-e^2)^\frac{1}{2}+O(\frac{1}{c})$ with $\nu =2\,\arctan(\sqrt{\frac{1+e}{1-e}}\,\tan\frac{u}{2})$ and $\alpha$ refers to proper initial value which will be defined later. Note that the given expression of Eq.(\ref{rhs}) is somewhat deceiving. Because the $\pm$ sign in front of $\nu$ dependent term changes as $\cos\kappa_1$ passes its maximum $x_3$ or minimum $x_2$. For example, if $\cos\kappa_1$ arrives at $x_3$ when $\nu=\nu_3$, then when $\nu>\nu_3$ it would be better if right hand side of Eq.(\ref{rhs}) is written as
\begin{align}\label{rhsex}
    \alpha + 2\,\frac{\nu_3+e\sin\nu_3}{c^2\,L^3}-\frac{\nu+e\sin\nu}{c^2\,L^3}\,,
\end{align}
hence it has a periodic structure behind and more complicated feature than presented in Eq.(\ref{rhs}). But as we can see later, the presented expression of Eq.(\ref{rhs}) will be justified.\\
\indent Now let us solve the left hand side of Eq.(\ref{ck1Eq}) keeping in mind that $x_2\leq x \leq x_3 < x_1$,
\begin{align}\label{lhs}
&(l.h.s)=\frac{dx}{\sqrt{A(x-x_1)(x-x_2)(x-x_3)}}\\\notag
= &\frac{dx}{\sqrt{A(x_1-x_2)}\sqrt{\frac{x_1-x}{x_1-x_2}}\sqrt{x-x_2}\sqrt{x_3-x_2}\sqrt{\frac{x_3-x}{x_3-x_2}}}\\\notag
=&\frac{\frac{1}{\sqrt{x_3-x_2}}\frac{1}{\sqrt{x-x_2}}dx}{\sqrt{A(x_1-x_2)}\sqrt{1-\frac{x_3-x_2}{x_1-x_2}\frac{x-x_2}{x_3-x_2}}\sqrt{1-\frac{x-x_2}{x_3-x_2}}}.
\end{align} 
Since $0\leq\frac{x-x_2}{x_3-x_2}\leq1$, we can introduce a new variable $y$ defined through 
\begin{align}\label{y}
\sin^2 y = \frac{x-x_2}{x_3-x_2}\,
\end{align}
which is restricted to $0\leq y\leq \frac{\pi}{2}$.
Then the last line of the Eq.(\ref{lhs}) becomes
\begin{align}\label{lhs2}
&\frac{\frac{1}{\sqrt{x_3-x_2}}\frac{1}{\sqrt{x-x_2}}dx}{\sqrt{A(x_1-x_2)}\sqrt{1-\frac{x_3-x_2}{x_1-x_2}\sin^2y}\,\cos y}\\\notag
=& \frac{ 2\,dy}{\sqrt{A(x_1-x_2)}\sqrt{1-\frac{x_3-x_2}{x_1-x_2}\sin^2y}}\,.
\end{align}
Identifying this integral as the Elliptic Integral of the first kind $F$,
\begin{align}
    F(\phi, \,k):=\int^\phi \frac{d\theta}{\sqrt{1-k^2\,\sin^2 \theta}}\,,
\end{align}
the Eq.(\ref{lhs2}) can be written in the following compact form
\begin{align}\label{ck1lhs}
\frac{2}{\sqrt{A(x_1-x_2)}}\,F\left(\arcsin\sqrt{\frac{\cos\kappa_1-x_2}{x_3-x_2}},\sqrt{\frac{x_3-x_2}{x_1-x_2}}\right)\,.
\end{align}
There could be other possible forms instead of Eq.(\ref{ck1lhs}) but, under the assumption of $x_2\leq\cos\kappa_1\leq x_3$, the above form is well defined. \\
\indent Finally by integrating both sides of Eq.(\ref{ck1Eq}), Keplerian parametrization of $\cos\kappa_1$ can be obtained as,
\begin{align}\label{ck1}
\cos\kappa_1\,&=\, x_2+(x_3-x_2)\,\text{sn}^2(\Upsilon\,,\,\beta)\,\\
&= x_2\,\text{cn}^2(\Upsilon\,,\,\beta)+x_3\,\text{dn}^2(\Upsilon\,,\,\beta)\,,\notag
\end{align}
 where $\Upsilon=\frac{\sqrt{A(x_1-x_2)}}{2} (\alpha +\frac{\nu+e\sin\nu}{c^2\,L^3})$, $\beta =\sqrt{\frac{x_3-x_2}{x_1-x_2}}$ and $\text{sn},\text{cn}$ and $\text{dn}$ are the sine, cosine and tangent like Jacobi elliptic functions \cite{E.T.WhittakerSc.D.2009}.\\
 \indent It is worth pointing out that $\pm$ sign which was supposed to be inside the definition of $\Upsilon$ according to Eq.(\ref{rhs}), is removed. Because of symmetric property, $\text{sn}^2(X,b)=\text{sn}^2(-X,b)$, we have a freedom to choose definition of $\Upsilon$ among four cases of  $\pm\frac{\sqrt{A(x_1-x_2)}}{2} (\alpha \pm\frac{\nu+e\sin\nu}{c^2\,L^3})$. Firstly, we have chosen $\frac{\sqrt{A(x_1-x_2)}}{2} (\pm\alpha +\frac{\nu+e\sin\nu}{c^2\,L^3})$ so that the sign of $\nu$ dependent term is positive. And we make $\pm$ sign in front of $\alpha$ be absorbed into the definition of $\alpha$. That is, 
 \begin{align}
 \alpha = \pm\frac{2}{\sqrt{A(x_1-x_2)}}\,F\left(\arcsin\sqrt{\frac{\cos\kappa_{1(0)}-x_2}{x_3-x_2}},\,\beta\right)\,,
 \end{align}
where $\cos\kappa_{1(0)}$ is an initial value of $\cos\kappa_1$ when $t=t_0$, or $u=0$. The $\pm$ sign is determined as : $+$ for the case that $\frac{d\cos\kappa_1}{dt}>0$ initially while $-$ for $\frac{d\cos\kappa_1}{dt}<0$ initially. Lastly, as we have seen, the expression of Eq.(\ref{rhs}) could not capture its oscillating feature. But we do not need to be concerned of it since the final form of our solution Eq.(\ref{ck1}) trivially captures the feature. Taking Eq.(\ref{rhsex}) as an example again, the reflection symmetry of sn, $\text{sn}(X+b,d)=\text{sn}(X-b,d)$ when sn has its maximum or minimum at $X$, allows (here we intentionally omit $\frac{\sqrt{A(x_1-x_2)}}{2}$ and $\beta$)
\begin{align}
    &\text{sn}\left(\alpha + 2\frac{\nu_3+e\sin\nu_3}{c^2\,L^3}-\frac{\nu+e\sin\nu}{c^2\,L^3}\right)\\\notag
    =\,&\text{sn}\left(\alpha +\frac{\nu+e\sin\nu}{c^2\,L^3}\right)
\end{align}
so that our solution is justified to be valid independent of duration.\\
\indent In the case of $A<0$ \textit{i.e.} $m_1<m_2$, the form of our solution is still valid with rearranging the order of the roots as $x_1<x_3\leq x_2$. Also in the case of no dynamics \textit{i.e.} $\frac{d\cos\kappa_1}{dt}=0$ all the time, our solution is valid with $x_2=x_3$. In the case of $A=0$ \textit{i.e.} $m_1=m_2$, our solution is still valid with $x_1\rightarrow\infty$ and thus $\beta=0$.\\
\indent Additionally, because the sine-like Jacobian elliptic function $\text{sn}(\Upsilon,\beta)$ is periodic, we can get the period of the precessional motion. Let $N$ be the number of orbital cycles. During $N$ cycles, $\Upsilon$ increases as
\begin{align}
\Delta\Upsilon_N= \frac{\sqrt{A(x_1-x_2)}\,\pi\,N}{c^2\,L^3}
\end{align}
while $\text{sn}^2(\Upsilon,\beta)$ has a following period \cite{E.T.WhittakerSc.D.2009},
\begin{align}
\Delta\Upsilon_P=\pi\,\mathcal{F}_{2,1}\left(\frac{1}{2},\frac{1}{2};1;\beta^2\right).
\end{align}
So $\Delta\Upsilon_P=\Delta\Upsilon_N$ yields the number of cycles $N=N_{\text{prec}}$ of the binary system during one period of precession,
\begin{align}
N_{\text{prec}} =   \mathcal{F}_{2,1}\left(\frac{1}{2},\frac{1}{2};1;\beta^2\right)\frac{c^2\,L^3}{\sqrt{A\,(x_1-x_2)}}\,
\end{align}
where $\mathcal{F}_{2,1}$ is a hypergeometric function.\\
Furthermore, invoking the definitions of $\sigma_1$ and $\sigma_2$, we can get the relationship between $\gamma$, $\kappa_1$ and $\kappa_2$
\begin{subequations}\label{cgck2}
\begin{align}
\cos\gamma &=\sigma_1 +\frac{L}{S_2}\frac{\delta_1-\delta_2}{\delta_2}\cos\kappa_1\,,\\
\cos\kappa_2 &= \sigma_2-\frac{\delta_1}{\delta_2}\frac{S_1}{S_2}\cos\kappa_1\,.
\end{align}
\end{subequations}
 \indent  Now, in the non-inertial frame where the orbital angular momentum is fixed as $\vek{L}= (0,0,L)$, we have a parametrization on how the spin angular momentums evolve around $\vek{L}$. \\
\indent In the non inertial frame, each angular momentum is expressed as, 
\begin{subequations}\label{amNF}
\begin{align}
\vek{J} = \Lambda^{-1,T}\left(\begin{array}{ccc}
0\\
0\\
J\\
\end{array}\right)_{i}=\left(\begin{array}{ccc}
0\\
J\,\sin\xi_2\\
J\,\cos\xi_2\\
\end{array}\right)_{n}\,,
\end{align}
and
\begin{align}\label{amN}
\vek{L}&=\left(\begin{array}{ccc}
0\\
0\\
L\\
\end{array}\right)_{n}\,,\\
\vek{S}_1&=\,S_1\left(\begin{array}{ccc}
\sin\kappa_1\cos\xi_3\\
\sin\kappa_1\sin\xi_3\\
\cos\kappa_1\\
\end{array}\right)_{n}\,,\\
\vek{S}_2&=\,S_2\left(\begin{array}{ccc}
\sin\kappa_2\cos(\xi_3+\Delta\psi)\\
\sin\kappa_2\sin(\xi_3+\Delta\psi)\\
\cos\kappa_2\\
\end{array}\right)_{n}\,,
\end{align}
\end{subequations}
where $ \Lambda^{-1,T}$ is the inverse matrix of the transpose of $\Lambda$ and the subscripts $n$ and $i$ refer to non-inertial and inertial frame respectively, where components of vector are described.  Since $\vek{J}=\vek{L}+\vek{S}_1+\vek{S}_2$ , we get the following three relations
\begin{subequations}\label{relations}
\begin{align}
0\,&=\,S_1\,\sin\kappa_1\,\cos\xi_3+S_2\,\sin\kappa_2\,\cos(\xi_3+\Delta\psi)\\
J\,\sin\xi_2\,&=\,\,S_1\,\sin\kappa_1\,\sin\xi_3+S_2\,\sin\kappa_2\,\sin(\xi_3+\Delta\psi)\\
J\,\cos\xi_2 &= L+S_1\cos\kappa_1+S_2\cos\kappa_2\,.\\\notag
\end{align}
\end{subequations}
Fom Eq.(\ref{relations}c), we get $\cos\xi_2$ in terms of $\cos\kappa_1$,
\begin{align}\label{xi2}
\cos\xi_2 
&= \frac{1}{J}(L+S_1\cos\kappa_1+S_2\cos\kappa_2)\,,\\\notag
&= \frac{L+S_2\sigma_2}{J}-\frac{S_1\,(\delta_1-\delta_2)}{J\delta_2}\cos\kappa_1\,.
\end{align}
On the other hands, from Eq.(\ref{relations}a) and Eq.(\ref{relations}b), we get the expression for $\xi_3$
\begin{align}
\left(\begin{array}{ccc}
\sin\xi_3\\
\cos\xi_3\\
\end{array}\right) = 
\left(\begin{array}{ccc}
\frac{S_1\,\sin\kappa_1+S_2\,\cos\Delta\psi\,\sin\kappa_2}{J\,\sin\xi_2}\\
\frac{S_2\,\sin\Delta\psi\,\sin\kappa_2}{J\,\sin\xi_2}\\
\end{array}\right)\,,
\end{align}
with Eq.(\ref{cosdeltapsi}),
\begin{align}
    \sin\xi_3=\frac{S_1\,\sin^2\kappa_1+S_2\,(\cos\gamma-\cos\kappa_1\,\cos\kappa_2)}{J\,\sin\kappa_1\,\sin\xi_2}.
\end{align}
This completes parametrization of all relative angles in Kepelerian type.\\

\subsection{The Keplerian-type Parametrization of the Orbital Angular Momentum}\label{solutionOAM}

\noindent Now, we are in the position to solve a Keplerian parametrization of $\xi_1$. 
The easiest way is to get components of $\frac{d\vek{L}}{dt}$ in the non inertial frame, and solve Eq.(\ref{lEq}), also in the non inertial frame.\\
\indent Starting from the inertial-expression of $\frac{d\vek{L}}{dt}$, we can get components of $\frac{d\vek{L}}{dt}$ in the non-inertial frame\,,
\begin{align}
\frac{d\vek{L}}{dt} &=L\,\frac{d\vek{k}}{dt}\\\notag
&=\,L\,\left(\begin{array}{ccc}
\cos\xi_1\,\sin\xi_2\,\dot{\xi}_1+\sin\xi_1\,\cos\xi_2\,\dot{\xi}_2\\
\sin\xi_1\,\sin\xi_2\,\dot{\xi}_1-\cos\xi_1\,\cos\xi_2\,\dot{\xi}_2\\
-\sin\xi_2\,\dot{\xi}_2\\
\end{array}\right)_{i}\\\notag
&=\,L\,\left(\begin{array}{ccc}
\sin\xi_2\,\dot{\xi}_1\\
-\dot{\xi}_2\\
0\\
 \end{array}\right)_{n}\,.
\end{align}
From the componential expression of Eq.(\ref{lEq}) in the non-inertial frame, we get
\begin{widetext}
\begin{align}\label{xi1xi2}
\left(\begin{array}{ccc}
\sin\xi_2\,\dot{\xi}_1\\
-\dot{\xi}_2\\
0\\
\end{array}\right)_{n} =\frac{1}{c^2\,r^3}\left(\begin{array}{ccc}
S_1\,\delta_1\,\sin\kappa_1\,\sin\xi_3+S_2\,\delta_2\,\sin\kappa_2\,\sin(\xi_3+\Delta\psi)\\
-S_1\,\delta_1\,\sin\kappa_1\,\cos\xi_3-S_2\,\delta_2\,\sin\kappa_2\,\cos(\xi_3+\Delta\psi)\\
0\\
\end{array}\right)_{n}\,.
\end{align}
The first row in the equation can be written in terms of $\cos\kappa_1$ using the relations Eqs.(\ref{cgck2}) and (\ref{relations}), 

\begin{align}
\dot{\xi}_1 &= \frac{J\,\delta_2}{c^2\,r^3}\frac{-\delta_2 \big\{\delta_2 \,\big[(L+\sigma_2 \,S_2)^2-J^2\big]+S_1^2 \,(\delta_2-\delta_1)+\sigma_1 \,S_1\, S_2 \,(\delta_2-\delta_1)\big\}+S_1 \,(\delta_1-\delta_2) \, \big[L\, (\delta_1+\delta_2)+\delta_2 \,\sigma_2\, S_2\big]\,\cos\kappa_1}{\delta_2^2 \,\big[J^2-(L+\sigma_2 S_2)^2\big]+2 \,\delta_2 \,S_1\, (\delta_1-\delta_2) \, (L+\sigma_2\, S_2)\,\cos\kappa_1-S_1^2 \,(\delta_1-\delta_2)^2 \,\cos ^2\kappa_1}\,\\\notag
& =: \frac{1}{c^2\,r^3}\,\left(\frac{\beta_1}{\cos\kappa_1+\alpha_1}-\frac{\beta_2}{\cos\kappa_1+\alpha_2}\right)\,,
\end{align}
\noindent where $\alpha_1$, $\alpha_2$, $\beta_1$ and $\beta_2$ are defined as follows
\begin{subequations}
\begin{align}
\alpha_1 &=- \frac{\delta_2\,(J+L+S_2\,\sigma_2)}{S_1\,(\delta_1-\delta_2)}\,,\\
\alpha_2 &=- \frac{\delta_2\,(-J+L+S_2\,\sigma_2)}{S_1\,(\delta_1-\delta_2)}\,,\\
\beta_1&=-\frac{L^2\,\delta_1+J^2\,\delta_2+S_1\,(\delta_1-\delta_2)(S_1+S_2\,\sigma_1)+L\,S_2\,\delta_1\,\sigma_2+J\,\big[L+(\delta_1+\delta_2)+S_2\,\sigma_2\,\delta_2\big]}{2\,S_1\,(\delta_1-\delta_2)}\,,\\
\beta_2&=-\frac{L^2\,\delta_1+J^2\,\delta_2+S_1\,(\delta_1-\delta_2)(S_1+S_2\,\sigma_1)+L\,S_2\,\delta_1\,\sigma_2-J\,\big[L+(\delta_1+\delta_2)+S_2\,\sigma_2\,\delta_2\big]}{2\,S_1\,(\delta_1-\delta_2)}\,.
\end{align}
\end{subequations}
 The integration can be written as, again with $x=\cos\kappa_1$, 
\begin{align}
\int d\xi_1 
&=\pm\,\int \frac{dx}{\sqrt{A\,(x-x_1)\,(x-x_2)\,(x-x_3)}}\,\left(\frac{\beta_1}{x+\alpha_1}-\frac{\beta_2}{x+\alpha_2}\right)\,.
\end{align}
At this point, we encounter similar issue discussed around Eq.(\ref{rhs}). The sign of $\frac{dx}{dt}$ oscillates during evolution and this makes the integration messy. To resolve this, we use $y$ introduced in Eq.(\ref{y}) again but imposing that $y$ increases as $t$ increases instead of restricting the range of $y$. (In fact, by comparing with Eq.(\ref{ck1}), $y=\text{am}(\Upsilon,\beta)$.) Then, the integration can be written without the $\pm$ sign as
\begin{align}
\int d\xi_1= \int\frac{ 2\,dy}{\sqrt{A(x_1-x_2)}\sqrt{1-\frac{x_3-x_2}{x_1-x_2}\sin^2y}}\,\left(\frac{\frac{\beta_1}{\alpha_1+x_2}}{1-\frac{x_2-x_3}{\alpha_1+x_2}\,\sin^2y}-\frac{\frac{\beta_2}{\alpha_2+x_2}}{1-\frac{x_2-x_3}{\alpha_2+x_2}\,\sin^2y}\right)\,.
\end{align}
The above integration is identified as the Elliptic integral of the third kind \cite{E.T.WhittakerSc.D.2009}, denoted as $\Pi$,
\begin{align}
\Pi\,(a,\, b, \,c) := \int^b_0 \frac{1}{\sqrt{1-c^2\,\sin^2 \theta}}\frac{d\theta}{1-a\,\sin^2\theta}\,.
\end{align}
Therefore
\begin{align}\label{xi1}
\xi_1-\xi_{1(0)}
&=\frac{2}{\sqrt{A\,(x_1-x_2)}}\,\Big(\frac{\,\beta_1\,\Pi\,(\frac{x_2-x_3}{\alpha_1+x_2},\,\text{am}(\Upsilon,\,\beta),\,\beta)}{\alpha_1+x_2}\,-\,\frac{\beta_2\,\Pi\,(\frac{x_2-x_3}{\alpha_2+x_2},\,\text{am}(\Upsilon,\,\beta),\,\beta)}{\alpha_2+x_2}\Big)\,,
\end{align}
where $\xi_{1(0)}$ is an initially determined parameter and $\text{am}$ denotes the Jacobi amplitude.

\end{widetext}

Up to now Keplerian parametrization for evolution of angular momentum has been derived in the inertial frame. Especially, the orbital angular  momentum $\vek{L}$ dynamics is totally determined by Eqs.(\ref{xi1}) and (\ref{xi2}) up to leading order precession. The last step is a parametrization of relative motion of binaries up to the leading order of spin-orbit interaction.

\subsection{Keplerian-type Parametrization of the relative motion of binaries}\label{solutionRP}
\noindent From the Hamiltonian Eq.(\ref{Hamiltonian}), the equation governing $\vek{r}$ is
\begin{align}
\dot{\vek{r}} = \frac{\partial H}{\partial \vek{p}}\, = \,\vek{p}+\frac{1}{c^2\,r^3}\,(\delta_1\,\vek{S}_1+\delta_2\,\vek{S}_2)\times\vek{r}
\,,
\end{align}
or,
\begin{align}
\vek{p}\,=\,\dot{\vek{r}}-\frac{1}{c^2\,r^3}\,(\delta_1\,\vek{S}_1+\delta_2\,\vek{S}_2)\times\vek{r}.
\end{align}
As can be seen from Fig.~\ref{fig:frames}, $\vek{r}$ can be written,
\begin{align}
\vek{r} = \left(\begin{array}{ccc}
r \cos\varphi\\
r \sin\varphi\\
0
\end{array}\right)_{n}\,.
\end{align}
To get an explicit expression for $\dot{\vek{r}}$, we must take time derivative of $\vek{r}$ in the inertial frame, and bring it back to the non-inertial frame using the Euler matrix,
\begin{align}\label{dotrNF}
\dot{\vek{r}} = \left(\begin{array}{ccc}
\dot{r} \,\cos\varphi-r \,\sin\varphi\, (\dot{\xi}_1 \,\cos \xi_2+\dot{\varphi})\\
\dot{r}\, \sin \varphi+r\, \cos \varphi \,(\dot{\xi}_1 \,\cos\xi_2+\dot{\varphi})\\
r \,(-\dot{\xi}_1 \,\sin\xi_2 \,\cos\varphi+\dot{\xi}_2 \,\sin\varphi)
\end{array}\right)_{n}\,.
\end{align}
\begin{widetext}
\noindent Now we have all ingredients (Eqs.(\ref{amNF}) and (\ref{dotrNF})) to express the vector identity $\vek{L}=\vek{r}\times\vek{p}$ in non-inertial frame components,
\begin{align}\label{oamNF}
\left(\begin{array}{ccc}
0 \\
0\\
L
\end{array}\right)_{n}=\vek{L}=\vek{r}\times\vek{p}=\left(\begin{array}{ccc}
-\frac{D}{r} \,\sin\varphi \\
\frac{D}{r}\,\cos\varphi\\
r^2\,\frac{d\varphi}{dt}+r^2\,\cos\xi_2\,\dot{\xi}_1-\frac{S1\,\delta_1\,\cos\kappa_1+S2\,\delta_2\,\cos\kappa_2}{c^2\,r}
\end{array}\right)_{n}\,,
\end{align}
where 
\begin{align}
D =&\, r^3\,(-\sin\varphi\,\dot{\xi}_2+\sin\xi_2\,\cos\varphi\,\dot{\xi}_1)\\\notag
&-\frac{\cos\varphi\,(S_1\delta_1\,\sin\kappa_1\sin\xi_3+S_2\,\delta_2\,\sin\kappa_2\,\sin(\xi_3+\Delta\psi))-\sin\varphi(S_1\delta_1\,\cos\xi_3\sin\kappa_1+S_2\,\delta_2\,\cos(\xi_3+\Delta\psi)\sin\kappa_2)}{c^2}\,.
\end{align}

\end{widetext}
 Note that $D$ trivially vanishes as Eq.(\ref{oamNF}) requires as Eq.(\ref{xi1xi2}) is satisfied.
\noindent The third component of $\vL$ reads to the following equation
\begin{align}\label{dotphiEq}
\frac{d\varphi}{dt}
&= \frac{L}{r^2}+\frac{1}{c^2\,r^3}\,(S_1\,\delta_1\,\cos\kappa_1+S_2\,\delta_2\,\cos\kappa_2)-\cos\xi_2\,\dot{\xi}_1\,,\\\notag
&= \frac{L}{r^2}\,(1-\frac{\delta_1+\delta_2}{c^2\,r})+\frac{1}{c^2\,r^3}\,(\frac{\beta_1}{\alpha_1+\cos\kappa_1}+\frac{\beta_2}{\alpha_2+\cos\kappa_1})\,.
\end{align}

\noindent For perturbatively consistent integration of $\frac{d\varphi}{dt}$, we need 1.5PN accurate $r$ expression. It is because $\frac{d\varphi}{dt}$ depends on $r$ from Newtonian order while we want to get $\varphi$ with an accuracy of 1.5PN. For a moment, we need to find the Keplerian parametrization of $r$. Using $\vek{p}^2=\dot{r}^2+\frac{L^2}{r^2}$ and $H = E$,

\begin{align}\label{dotrEq}
\dot{r}^{\,2} = 2\,E +\frac{2}{r}-\frac{L^2}{r^2}-\frac{2\,(\vek{L}\cdot\vek{S}_\text{eff})}{c^2\,r^3}\,.
\end{align}
Note that $\vek{L}\cdot\vek{S}_\text{eff}$ is a constant with the following simplification
\begin{align}
\vek{L}\cdot\vek{S}_\text{eff} 
& = \vek{L}\cdot(\delta_1\,\vek{S}_1+\delta_2\,\vek{S}_2)\\\notag
& =\delta_1\,L\,S_1\,\cos\kappa_1+\delta_2\,L\,S_2\,\cos\kappa_2\\\notag
&= \delta_2\,L\,S_2\,\sigma_2\,.
\end{align}
Eq.(\ref{dotrEq}) is easily integrated \cite{Damour1985} and the radial motion is described, up to leading order of spin-orbit coupling, as
\begin{subequations}\label{radius}
\begin{align}
r & = a_r\,(1-e_r\,\cos u)\,,\\
n\,(t-t_0) &= u-e_t\,\sin u\,,
\end{align}
where 

\begin{align}
a_r &= -\frac{1}{2\,E}(1-\frac{2\,\delta_2\,S_2\,\sigma_2}{L}\,\frac{E}{c^2})\,,\\
n&= (-2\,E)^{3/2}\,,\\
e_r^2 &= 1+2\,E\,L^2+8\,(1+E\,L^2)\,\frac{\delta_2\,S_2\,\sigma_2}{L}\,\frac{E}{c^2}\,,\\
e_t^2 &= 1+2\,E\,L^2+\frac{4\,\delta_2\,S_2\,\sigma_2}{L}\,\frac{E}{c^2}\,.
\end{align}
\end{subequations}
Now let us go back to Eq.(\ref{dotphiEq}). We've already known that the integration of the second term of right hand side of Eq.(\ref{dotphiEq}) is the Elliptic integral of the thrid kind and the integration of the first term is easily done via Eq.(\ref{radius}),
\begin{widetext}
\begin{subequations}\label{varphi}
\begin{align}
\varphi-\varphi_0
& = (1+k)\,\bar{\nu}+\frac{2\,\beta_1\,\Pi(\frac{x_2-x_3}{\alpha_1+x_2},\,\text{am}(\Upsilon, \beta),\,\beta)}{\sqrt{A\,(x_1-x_2)}(\alpha_1+x_2)}\,+\frac{2\,\beta_2\,\Pi(\frac{x_2-x_3}{\alpha_2+x_2},\,\text{am}(\Upsilon,\beta),\,\beta)}{\sqrt{A\,(x_1-x_2)}(\alpha_2+x_2)}\,,
\end{align}
where
\begin{align}
k&=\frac{-1}{c^2\,L^2}\,\left(\delta_1+\delta_2+3\frac{\delta_2\,S_2\,\sigma_2}{L}\right)\,,\\
\bar{\nu} & =2\,\arctan\left(\sqrt{\frac{1+e_\varphi}{1-e_\varphi}}\,\tan\frac{u}{2}\right)\,,\\
e_\varphi^2 &= 1+2\,E\,L^2+4\,(\delta_1+\delta_2)\,( 1+2\,E\,L^2)\,\frac{E}{c^2}+4\,(3+4\,E\,L^2)\,\frac{\delta_2\,S_2\,\sigma_2}{L}\frac{E}{c^2}\,,
\end{align}
\end{subequations}
and $\varphi_0$ is an initially given value.
\end{widetext}
\indent In principle, we have all parametrizations for describing the Newtonian orbital motion with the leading order of spin-orbit coupling. In the inertial frame, it is written as
\begin{align}\label{radiusIF}
    \vek{r}=r\,\left(\begin{array}{ccc}
\cos\xi_1\,\cos\varphi-\sin\xi_1\,\cos\xi_2\,\sin\varphi\\
\sin\xi_1\,\cos\varphi+\cos\xi_1\,\cos\xi_2\,\sin\varphi\\
\sin\varphi\,\sin\xi_2
\end{array}\right)_{i}.
\end{align}


\section{Almost equal mass Approximation}\label{almostequalmass}
In this section, we restrict our parametrizations in the case of $\delta_2-\delta_1\sim0$, and provide the simpler version trucated at the order of $O((\delta_2-\delta_1)^2)$. Such a limiting case was considered by \cite{Tessmer2009} but the result was not provided in a closed form. Here we provide it only by elementary functions.  In the exactly equal mass case, the solution was derived by \cite{Konigsdorffer2005} analytically so that we have a chance to compare directly.

\indent For the purpose of expansion for nearly same mass binaries, we choose a dimensionless parameter $\delta$,
\begin{align}
\delta:= \delta_2-\delta_1 =\frac{3}{2}\,\frac{m_1-m_2}{M}
\end{align}
which is assumed to be non-negative (\textit{i.e.} $m_1\geq m_2$). The targeted approximant is achieved straightforwardly by changing of variable  $\delta_2\rightarrow\delta+\delta_1$ and expanding it in terms of $\delta$ up to the first order. We should be careful about the fact that, since $A\,(x-x_1)(x-x_2)(x-x_3)$ becomes the 2nd order polynomial in the case of equal mass, $x_1$ should blow up as $\delta\rightarrow0$ and indeed by concrete calculation, it turns out that $x_1=\frac{1}{\delta}\,\frac{\delta_1\,\left(S_1^2+2 \sigma_1 \,S_1 \,S_2+S_2^2\right)}{2\, L\, S_1}+O(\delta^0)$. And we further find that
\begin{widetext}
\begin{align}
    \frac{x_2-x_3}{\alpha_1+x_2}
   & =-\delta\,\frac{2 \,S_1\,S_2 \,\sqrt{\left(1-\sigma_1^2\right) \left(S_1^2+2 \sigma_1 S_1 S_2+\left(1-\sigma_2^2\right) S_2^2\right)}}{\delta_1\, \left(S_1^2+2 \sigma_1 S_1 S_2+S_2^2\right) (J+L+\sigma_2 S_2)}+O(\delta^2)\,,\\
        \frac{x_2-x_3}{\alpha_2+x_2}
    &=-\delta\,\frac{2\,  S_1 \,S_2\, \sqrt{\left(1-\sigma_1^2\right) \left(S_1^2+2 \sigma_1 S_1 S_2+\left(1-\sigma_2^2\right) S_2^2\right)}}{\delta_1\, \left(S_1^2+2 \sigma_1 S_1 S_2+S_2^2\right) (-J+L+\sigma_2 S_2)}+O(\delta^2)\,,\\
    \beta&=\delta\,\frac{4 \, L \,S_1\, S_2\, \sqrt{\left(1-\sigma_1^2\right) \,\left(S_1^2+2 \sigma_1 S_1 S_2+\left(1-\sigma_2^2\right) S_2^2\right)}}{{\delta_1\, \left(S_1^2+2 \sigma_1 S_1 S_2+S_2^2\right)^2}}+O(\delta^2)\,.
\end{align}
In order to get the Keplerian parametrizations for $\xi_1$ and $\varphi$ expanded up to the first order in $\delta$, it is essential to get the first order approximant of the type $\Pi(\mathcal{A},\,\text{am}(\Upsilon,\beta),\,\beta)$. With $\mathcal{A}\sim O(\delta^1)$, $\beta\sim O(\delta^1)$ and $\Upsilon\sim O(\delta^0)$, we have
\begin{align}
    \Pi(\mathcal{A},\,\text{am}(\Upsilon,\beta),\,\beta) =\left(1+\frac{\mathcal{A}}{2}\right)\,\Upsilon-\frac{\mathcal{A}}{4}\,\sin2\Upsilon+O(\delta^2)\,.
\end{align}

\noindent Then it staightforwardly follows that
\begin{align}
\xi_1-\xi_{1(0)}=\Xi_1\,\hat\Upsilon+\,\Xi_2\,\sin2\hat\Upsilon+O(\delta^2)\,,
\end{align}
where
\begin{align}
\Xi_1 &= \frac{2 J}{\sqrt{S_1^2+2 \sigma_1 S_1 S_2+S_2^2}}+\delta\,\frac{2 \, J\, S_2\, (\sigma_1 S_1+S_2)}{\delta_1 \left(S_1^2+2 \sigma_1 \,S_1\, S_2+S_2^2\right)^{3/2}}\,,\\
\Xi_2 &= \,-\delta\,\frac{  J\, \sqrt{1-\sigma_1^2} \,S_1\,S_2^2 \,\sigma_2 }{\delta_1 \,(S_1^2+2 \sigma_1\, S_1\, S_2+S_2^2)^{3/2} \sqrt{S_1^2+2 \,\sigma_1\, S_1\, S_2+(1-\sigma_2^2) \,S_2^2}}\,,
\end{align}
and $\hat\Upsilon$ is the truncated one,
\begin{align}
\hat\Upsilon= \alpha+\frac{\delta_1\,\sqrt{S_1^2+S_2^2+2\,S_1\,S_2\,\sigma_1}}{2\,c^2\, L^3}\,(\nu+e\,\sin\nu).
\end{align}
Similarly, Eq.(\ref{varphi}a) is expanded to
\begin{align}
\varphi-\varphi_0 = (1+\hat k)\,\hat\nu+\Phi_1\,\hat{\Upsilon}\,+\,\Phi_2\,\sin2\hat\Upsilon+O(\delta^2)\,,
\end{align}
where
\begin{subequations}
\begin{align}
\Phi_1 &= \frac{2 L}{\sqrt{S_1^2+2 \sigma_1 S_1 S_2+S_2^2}}+\delta\,\frac{2 \,  L\, S_1 \,(S_1+\sigma_1 S_2)}{\delta_1\, \left(S_1^2+2 \sigma_1 S_1 S_2+S_2^2\right)^{3/2}}\,,\\
\Phi_2&= \,\delta\,\frac{  (J^2-L^2-L\,S_2\,\sigma_2)\, \sqrt{1-\sigma_1^2} \,S_1 \,S_2}{\delta_1 \,(S_1^2+2 \sigma_1\, S_1\, S_2+S_2^2)^{3/2} \sqrt{S_1^2+2 \,\sigma_1\, S_1\, S_2+(1-\sigma_2^2) \,S_2^2}}\,,\\
\hat{k}\,&= -\frac{\delta_1 \,(2\, L+3 \,\sigma_2\, S_2)}{c^2 L^3}+\delta\,\frac{(-L-3\, \sigma_2\, S_2)}{c^2 L^3}\,,\\
\hat{e}^{\,2}&= 1+2 \,E\, L^2+\frac{4 \delta_1 \,E\, \left(4 \,E\, L^3+4 \,E\, L^2 \sigma_2 S_2+2 L+3 \sigma_2 S_2\right)}{c^2 L}+\delta\,\frac{4  \,E\, \left(2 \,E\, L^3+4 \,E\, L^2 \sigma_2 S_2+L+3 \sigma_2 S_2\right)}{c^2 L}\,,
\end{align}
\end{subequations}
with $\hat\nu=2\,\arctan\left(\sqrt{\frac{1+\hat{e}}{1-\hat{e}}}\,\tan\frac{u}{2}\right)$. Note that $\hat{k}$, $\hat{e}$ are just approximants of $k$ and $e_\varphi$ in Eqs.(\ref{varphi}) truncated at the order of $O(\delta^2)$.
\end{widetext}
\subsection{Equal mass binaries limit}
\indent From this we can easily compare the leading terms with the results of \cite{Konigsdorffer2005}, which deals with equal mass case. By setting $\delta=0$, the expression for $\xi_1$ becomes,
\begin{align}
\xi_1-\xi_{1(0)}= \frac{J\,\delta_1}{c^2\,L^3}\,(\nu+e\,\sin\nu)\,.
\end{align}
Similarly,
\begin{align}
\varphi-\varphi_0 = &\left(1-\frac{\delta_1\,(2L+3\,S_2\,\sigma_2)}{c^2\,L^3}\right)\,\hat\nu\,|_{\delta=0}\\\notag
&+\frac{\delta_1}{c^2\,L^2}\,(\nu+e\,\sin\nu)\,.
\end{align}
Since the followings hold perturbatively,
\begin{align}
    &2\arctan\left(a\,(1+b)\right)\\\notag
    =&2\arctan\left(a\right)+b\,\sin\left(2\arctan\left(a\right)\right)+O(b^2)\,,
\end{align}
and
\begin{align}
   &\sqrt{\frac{1+(a+b)}{1-(a+b)}} \\\notag
   =&\sqrt{\frac{1+a}{1-a}}\,\left(1+\frac{b}{1-a^2}\right)+O(b^2)
\end{align}
we can make some part of $\varphi$ more compact by introducing
\begin{align}
    \bar\nu'&:=\hat\nu\,|_{\delta=0}+\frac{\delta_1}{c^2\,L^2}\,e\,\sin\nu \\\notag
    &=\, 2\,\arctan\left(\sqrt{\frac{1+\hat{e}}{1-\hat{e}}}\,\tan\frac{u}{2}\,(1+\frac{\delta_1}{c^2\,L^2}e)\right)+O(\frac{1}{c^4})\,,\\\notag
    &=:\,2\,\arctan\left(\sqrt{\frac{1+e'}{1-e'}}\,\tan\frac{u}{2}\right)+O(\frac{1}{c^4})\,.
\end{align}
Now we arrive the compact form in the accuracy of $O(\frac{1}{c^4})$ such that
\begin{align}
\varphi-\varphi_0 
& = (1+k')\,\bar\nu'\,+O(\frac{1}{c^4}),
\end{align}
where
\begin{subequations}
\begin{align}
k'=&\, -\frac{\delta_1}{c^2\,L^2}\,(1+3\frac{S_2\,\sigma_2}{L})\,,\\
\bar\nu' =& \,2\,\arctan\left({\sqrt{\frac{1+e'}{1-e'}}\,\tan\frac{u}{2}}\right)\,,\\
e'^{\,2}=&\,1+2\,E\,L^2+4\,\delta_1\,(1+2\,E\,L^2)\frac{E}{c^2}\\\notag
&+4\,(3+4\,E\,L^2)\frac{\delta_1\,S_2\,\sigma_2}{L}\,\frac{E}{c^2}\,.
\end{align}
\end{subequations}
These results are coincident with the results of \cite{Konigsdorffer2005}.

\section{Conclusion}
We have solved the spin precession equation and the Newtonian orbital motion in ADM gauge for binary system with arbitrary spins, mass ratio and eccentricity up to leading order of spin orbit coupling. We arrived at quasi-Keplerian parametric solutions in a simple closed form. The elliptic functions are essential in our parametrizations and they have been thoroughly speculated by mathematicians. So our solutions are expected to give us systematic and mathematically deep understandings on how rotating bodies move in gravitational field. On the other hand, numerical computations for the elliptic functions are rather expensive compared to the elementary functions. However, our rough numercial estimations suggest that a couple of the constants defined in this paper such as $\beta$, $\frac{x_2-x_3}{\alpha_1+x_2}$ tend to be very small in almost all initial configurations, implying that we would be able to re-express the elliptic functions in terms of the elementary functions without significant numercial errors. This could be our future work.\\
\indent We also expect that the solutions presented here can be directly used to get efficient ready-to-use time domain gravitational waveform templates modulated by spin precession and swings of orbital plane, which would be useful in analyzing the gravitational wave data. Furthermore, the result of this paper will provide useful knowledges to complete phenomenological models such as \textit{effective-one-body} which requires both numerical calibrations and analytic solutions. \\
\section*{Acknowledgement}
Gihyuk Cho acknowledges financial support through BK21 plus program of the Ministry of Education, Government of korea.
\newpage
\bibliographystyle{apsrev}
\bibliography{GihyukCho.bib}

\end{document}